\documentclass{article}
\usepackage{spconf,amsmath,graphicx,pgfplots,tikz}
\usepackage{xfrac,bm,amssymb,multirow,algorithm,algorithmic}
\usepackage[caption=false]{subfig}
\pgfplotsset{compat=1.14}

\title{SVD-PHAT: A Fast Sound Source Localization Method}
\name{Fran\c{c}ois Grondin, James Glass\thanks{This work was supported in part by the Toyota Research Institute and by the Fonds de recherche du Qu\'{e}bec –- Nature et technologies.}}
\address{Computer Science and Artificial Intelligence Laboratory\\
Massachusetts Institute of Technology\\
    Cambridge, MA 02139, USA \\
    \small\texttt{\{fgrondin,glass\}@mit.edu}}

\DeclareMathOperator*{\argmax}{arg\,max}

\DeclareMathOperator{\Tr}{Tr}
\DeclareMathOperator{\atantwo}{atan2}
\newcommand{\xMapsto}[2][]{\ext@arrow 0599{\Mapstofill@}{#1}{#2}}

\newcommand{\aiMiCx}{0}
\newcommand{\aiMiCy}{0}
\newcommand{\aiMiCz}{-5.0}
\newcommand{\aiMiiCx}{0}
\newcommand{\aiMiiCy}{0}
\newcommand{\aiMiiCz}{-3.3}
\newcommand{\aiMiiiCx}{0}
\newcommand{\aiMiiiCy}{0}
\newcommand{\aiMiiiCz}{-1.7}
\newcommand{\aiMivCx}{0}
\newcommand{\aiMivCy}{0}
\newcommand{\aiMivCz}{0}
\newcommand{\aiMvCx}{0}
\newcommand{\aiMvCy}{0}
\newcommand{\aiMvCz}{1.7}
\newcommand{\aiMviCx}{0}
\newcommand{\aiMviCy}{0}
\newcommand{\aiMviCz}{3.3}
\newcommand{\aiMviiCx}{0}
\newcommand{\aiMviiCy}{0}
\newcommand{\aiMviiCz}{5.0}

\newcommand{\aiiMiCx}{0}
\newcommand{\aiiMiCy}{0}
\newcommand{\aiiMiCz}{0}
\newcommand{\aiiMiiCx}{5}
\newcommand{\aiiMiiCy}{0}
\newcommand{\aiiMiiCz}{0}
\newcommand{\aiiMiiiCx}{2.5}
\newcommand{\aiiMiiiCy}{4.3}
\newcommand{\aiiMiiiCz}{0}
\newcommand{\aiiMivCx}{-2.5}
\newcommand{\aiiMivCy}{4.3}
\newcommand{\aiiMivCz}{0}
\newcommand{\aiiMvCx}{-5.0}
\newcommand{\aiiMvCy}{0}
\newcommand{\aiiMvCz}{0}
\newcommand{\aiiMviCx}{-2.5}
\newcommand{\aiiMviCy}{-4.3}
\newcommand{\aiiMviCz}{0}
\newcommand{\aiiMviiCx}{2.5}
\newcommand{\aiiMviiCy}{-4.3}
\newcommand{\aiiMviiCz}{0}
\newcommand{\aiiiMiCx}{0}
\newcommand{\aiiiMiCy}{0}
\newcommand{\aiiiMiCz}{0}
\newcommand{\aiiiMiiCx}{-5}
\newcommand{\aiiiMiiCy}{0}
\newcommand{\aiiiMiiCz}{0}
\newcommand{\aiiiMiiiCx}{5}
\newcommand{\aiiiMiiiCy}{0}
\newcommand{\aiiiMiiiCz}{0}
\newcommand{\aiiiMivCx}{0}
\newcommand{\aiiiMivCy}{-5}
\newcommand{\aiiiMivCz}{0}
\newcommand{\aiiiMvCx}{0}
\newcommand{\aiiiMvCy}{5}
\newcommand{\aiiiMvCz}{0}
\newcommand{\aiiiMviCx}{0}
\newcommand{\aiiiMviCy}{0}
\newcommand{\aiiiMviCz}{-5}
\newcommand{\aiiiMviiCx}{0}
\newcommand{\aiiiMviiCy}{0}
\newcommand{\aiiiMviiCz}{5}

\begin{document}
%
\maketitle
\begin{abstract}
This paper introduces a new localization method called SVD-PHAT.
The SVD-PHAT method relies on Singular Value Decomposition of the SRP-PHAT projection matrix.
A k-d tree is also proposed to speed up the search for the most likely direction of arrival of sound.
We show that this method performs as accurately as SRP-PHAT, while reducing significantly the amount of computation required.

\end{abstract}

\section{Introduction}
\label{sec:introduction}

Distant speech processing is a challenging task, as the target sound source is usually corrupted by noise from the environment and is degraded by reverberation \cite{tang2018study}.
Beamforming methods are often used as a preprocessing step to enhance the corrupted speech signal using multiple microphones.
Many beamforming methods, such as Delay and Sum (DS), Geometric Source Separation (GSS) \cite{parra2001geometric} and Minimum Variance Distortionless Response (MVDR) \cite{habets2010new}, require the target source direction of arrival (DOA).
Sound source localization consists in estimating this DOA, and often relies on Multiple Signal Classification (MUSIC) \cite{schmidt1986multiple} or Steered-Response Power Phase Transform (SRP-PHAT) \cite{dibiase2001robust} methods.

MUSIC is based on Standard Eigenvalue Decomposition (SEVD-MUSIC), and was initially used for narrowband signals, then adapted to broadband signals to make localization robust to additive noise \cite{ishi2009evaluation}.
The latter method however assumes that the target signal is more powerful than noise.
To cope with this limitation, MUSIC based on Generalized Eigenvalue Decomposition (GEVD-MUSIC) handles scenarios when noise is more powerful than the signal of interest \cite{nakamura2011intelligent}.
Alternatively, MUSIC based on Generalized Singular Value Decomposition (SVD-MUSIC), reduces the computational load of GEVD-MUSIC and improves DOA estimation accuracy \cite{nakamura2012real}.
However, all MUSIC-based methods require performing online eigenvalue or singular value decompositions, which involve a significant amount of computations, and make real-time processing more challenging.

SRP-PHAT computes the Generalized Cross-Correlation with Phase Transform (GCC-PHAT) between each pair of microphones \cite{brandstein1997robust}.
The exact SRP-PHAT solution involves fractional Time-Difference of Arrival (TDOA) delays, and requires a significant amount of computation.
The Fast Fourier Transform (FFT) is thus often used to speed up the computation of GCC-PHAT, which makes this method appealing for real-time applications \cite{grondin2013manyears,valin2007robust}.
However, using the FFT restricts the transform to discrete TDOA values, which reduces localization accuracy.
Interpolation \cite{jacovitti1993discrete,mccormick2013approach,viola2005spline}, fractional delay estimation \cite{maskell1999estimation} and fractional Fourier transform \cite{sharma2007time} attempt to overcome the FFT discretization drawback.
Moreover, searching for sound source involves a significant amount of computations when scanning the 3D-space.
Stochastic region contraction \cite{do2007real}, hierarchical search \cite{zotkin2004accelerated, do2009stochastic, nunes2014steered} and vectorization \cite{lee2010vectorized} are proposed to speed up scanning, but are usually restricted to a 2D surface.

In this paper, we propose a new method inspired from the original SRP-PHAT approach, called SVD-PHAT.
The objective is to reduce the amount of computations typically involved in the exact SRP-PHAT, while preserving its accuracy.
The proposed technique relies on SVD to generate a transform related to the matrix geometry that maps the initial observations to a smaller subspace.
This subspace is then searched with a k-d tree, which returns the estimated DOA.

\section{SRP-PHAT}
\label{sec:srpphat}

SRP-PHAT relies on the TDOA estimation for all pairs of microphones (for an array with $M$ microphones, there are $P = M(M-1)/2$ possible pairs).
The TDOA (in sec) corresponds to the difference between the distance from the source $\mathbf{s}_q \in \mathbb{R}^{3}$ to microphone $i$ at position $\mathbf{r}_i \in \mathbb{R}^{3}$, and the distance between the same source and another microphone $j$ at position $\mathbf{r}_j \in \mathbb{R}^{3}$, divided by the speed of sound in air $c \in \mathbb{R}^+$ (in m/sec).
Since all signals are discretized in time, it is also convenient to express the TDOA in terms of samples by adding the sample rate ($f_S \in \mathbb{R}^+$) in the expression, as shown in (\ref{eq:srpphat_tdoa}).
\begin{equation}
    \tau_{q,i,j} = \frac{f_S}{c}\left(\lVert \mathbf{s}_q - \mathbf{r}_i \rVert_2 - \lVert \mathbf{s}_q - \mathbf{r}_j \rVert_2\right)
    \label{eq:srpphat_tdoa}
\end{equation}
where $\lVert\dots\rVert_2$ stands for the Euclidean norm.

In most microphone array configurations, the array aperture is small compared to the distance between the source and the array, such that the farfield assumption holds.
In this case, (\ref{eq:srpphat_tdoa}) can be formulated as in (\ref{eq:srpphat_farfield}).
\begin{equation}
    \tau_{q,i,j} = \frac{f_S}{c}\left(\mathbf{r}_j-\mathbf{r}_i\right)\cdot \frac{\mathbf{s}_q}{\lVert\mathbf{s}_q\rVert_2}
    \label{eq:srpphat_farfield}
\end{equation}

Let $x_m[n]$ be the signal of microphone $m$ in the time domain.
The expression $X^l_{m}[k] \in \mathbb{C}$ is obtained with a Short Time Fourier Transform (STFT) with a sine window, where $N \in \mathbb{N}$ and $\Delta N \in \mathbb{N}$ stand for the frame and hop sizes in samples, respectively, and $k \in \mathbb{N} \cap [0,N/2]$ and $l \in \mathbb{N}$ stand for the frequency bin and frame indexes, respectively.
For clarity, the frame index $l$ is omitted in this paper without loss of generality.
The normalized cross-spectrum for each pair of microphones $(i,j)$ (where $i \neq j$) corresponds to the expression $X_{i,j}[k] \in \mathbb{C}$ in (\ref{eq:srpphat_phat}).
The operators $\{\dots\}^*$ and $|\dots|$ stand for the complex conjugate and the absolute value, respectively.
\begin{equation}
    X_{i,j}[k] = \frac{X_{i}[k]X_{j}[k]^*}{|X_{i}[k]||X_{j}[k]|}
    \label{eq:srpphat_phat}
\end{equation}

In the frequency domain, the TDOA $\tau_{q,i,j}$ leads to the coefficient $W_{q,i,j}[k] \in \mathbb{C}$ in (\ref{eq:srpphat_W}) according to SRP-PHAT beamforming.
\begin{equation}
    W_{q,i,j}[k] = \exp{\left(2\pi\sqrt{-1} k\tau_{q,i,j}/N\right)}
    \label{eq:srpphat_W}
\end{equation}

For each potential source position located at $\mathbf{s}_q$, SRP-PHAT returns an energy value expressed by $Y_q \in \mathbb{R}$, where $\Re\{\dots\}$ extracts the real part.
\begin{equation}
    Y_q = \Re\left\{\sum_{i=1}^{M}{\sum_{j=(i+1)}^{M}{\sum_{k=0}^{N/2}{W_{q,i,j}[k]X_{i,j}[k]}}}\right\}
    \label{eq:srpphat_Yq}
\end{equation}

The estimated direction of arrival (DOA) of sound corresponds to the position denoted by $\mathbf{s}_{\bar{q}}$, where $\bar{q}$ is obtained in (\ref{eq:srpphat_qmax}).
Moreover, the scalar $Y_{\bar{q}}$ is often used to discriminate a valid sound source from backgroud noise.
\begin{equation}
\bar{q} = \argmax_q{\{Y_q\}}
\label{eq:srpphat_qmax}
\end{equation}

Computing $Y_q$ for $q \in \mathbb{N} \cap [1, Q]$ as in (\ref{eq:srpphat_qmax}) involves a complexity order of $\mathcal{O}(QPN)$, and searching for the best potential source results in (\ref{eq:srpphat_qmax}) leads to a $\mathcal{O}(Q)$ search.
When the number of points to scan ($Q$) gets large, the SRP-PHAT involves numerous computations, which makes the method less suitable for real-time applications.
The proposed SVD-PHAT method described in the next section aims to alleviate this limitation.

\section{SVD-PHAT}
\label{sec:svdphat}

To define the SVD-PHAT method, it is convenient to start from SRP-PHAT expressed in matrix form.
We define the vector $\mathbf{X} \in \mathbb{C}^{P(N/2+1) \times 1}$ in (\ref{eq:svdphat_Xv}), which concatenates all normalized cross-spectra previously introduced in (\ref{eq:srpphat_phat}).
\begin{equation}
    \mathbf{X} = \left[
        \begin{array}{cccc}
            X_{1,2}[0] & X_{1,2}[1] & \cdots & X_{M-1,M}[N/2] \\
        \end{array}
    \right]^T
    \label{eq:svdphat_Xv}
\end{equation}

Similarly, the matrix $\mathbf{W} \in \mathbb{C}^{Q \times P(N/2+1)}$ holds all the SRP-PHAT coefficients:
\begin{equation}
\mathbf{W} = \left[
\begin{array}{cccc}
    W_{1,1,2}[0] & W_{1,1,2}[1] & \cdots & W_{1,M-1,M}[N/2] \\
    \vdots & \vdots & \ddots & \vdots \\
    W_{Q,1,2}[0] & W_{Q,1,2}[1] & \cdots & W_{Q,M-1,M}[N/2] \\
\end{array}
\right]
\label{eq:svdphat_Wm}
\end{equation}

Finally, the vector $\mathbf{Y} \in \mathbb{C}^{Q \times 1}$ stores the SRP-PHAT energy for all $Q$ potential sources and is obtained from the following matrix multiplication:
\begin{equation}
    \mathbf{Y} = \left[
        \begin{array}{ccc}
            Y_1 & \dots & Y_Q
        \end{array}
    \right]^T = \Re\{\mathbf{W}\mathbf{X}\}
    \label{eq:svdphat_Yv}
\end{equation}

As mentionned for SRP-PHAT, this matrix multiplication is computationally expensive when there are numerous potential source positions to scan.
To cope with this limitation, we propose to perform Singular Value Decomposition on the matrix $\mathbf{W}$, where $\mathbf{U} \in \mathbb{C}^{Q \times K}$, $\mathbf{S} \in \mathbb{C}^{K \times K}$ and $\mathbf{V} \in \mathbb{C}^{P(N/2+1) \times K}$, as shown in (\ref{eq:svdphat_svd}).
\begin{equation}
    \mathbf{W} \approx  \mathbf{U}\mathbf{S}\mathbf{V}^H
    \label{eq:svdphat_svd}
\end{equation}
where $\{\dots\}^H$ stands for the Hermitian operator.

The parameter $K \in \mathbb{N}\ \cap\ ]0,K_{max}]$, where the upper bound $K_{max} = \max\{Q,P(N/2+1)\}$, is chosen to ensure accurate reconstruction of $\mathbf{W}$, according to the condition in (\ref{eq:svdphat_trace}), where the user-defined parameter $\delta$ is a small positive value that models the tolerable reconstruction error.
The operator $\Tr\{\dots\}$ stands for the trace of the matrix.
\begin{equation}
    \Tr{\{\mathbf{S}\mathbf{S}^T\}} \geq (1 - \delta)\Tr{\{\mathbf{W}\mathbf{W}^H\}}
    \label{eq:svdphat_trace}
\end{equation}

The vector $\mathbf{Z} \in \mathbb{C}^{K \times 1}$ results from the projection of the observations $\mathbf{X}$ in the $K$-dimensions subspace:
\begin{equation}
    \mathbf{Z} = \mathbf{V}^H\mathbf{X}
    \label{eq:svdphat_Zv}
\end{equation}

The matrix $\mathbf{D} \in \mathbb{C}^{Q \times K}$ is obtained in (\ref{eq:svdphat_Dv}) and can be decomposed in a set of $Q$ vectors $\mathbf{D}_q \in \mathbb{C}^{1 \times K}$:
\begin{equation}
    \mathbf{D} = \mathbf{U}\mathbf{S} = \left[
    \begin{array}{cccc}
        \mathbf{D}_1^T & \mathbf{D}_2^T & \dots & \mathbf{D}_Q^T \\
    \end{array}
    \right]^T
    \label{eq:svdphat_Dv}
\end{equation}

The index of the most likely DOA obtained in (\ref{eq:srpphat_qmax}) now corresponds to:
\begin{equation}
    \bar{q} = \argmax_{q}{\{\Re\{\mathbf{D}_{q}\cdot\mathbf{Z}^H\}\}}
    \label{eq:svdphat_qmax2}
\end{equation}

One way to find the correct value of $q$ in (\ref{eq:svdphat_qmax2}) consists in computing every $Y_q$ for $q \in \mathbb{N} \cap [1, Q]$, and then finding the index $q$ that leads to the maximum value, which obviously involves a significant amount of computations, as the complexity order is linear ($\mathcal{O}(Q)$).
It is therefore relevant to look for an alternate cost function that would allow a more efficient search.
For all values of $q$, the expressions $\lVert \mathbf{D}_q \rVert_2$ are almost identical (when the reconstruction meets condition in (\ref{eq:svdphat_trace}) for a small value of $\delta$) , but do not necessary equal to 1.
We thus define the new vectors $\hat{\mathbf{D}}_q = \mathbf{D}_q / \lVert \mathbf{D}_q\rVert_2$ and the normalized vector $\hat{\mathbf{Z}} = \mathbf{Z} / \lVert \mathbf{Z} \rVert_2$.
With $\lVert\hat{\mathbf{D}}_q\rVert^2_2 = 1$ and $\lVert\hat{\mathbf{Z}}\rVert^2_2 = 1$, the dot product can therefore be expressed as follows: 
\begin{equation}
    \Re\{\mathbf{D}_q \cdot \mathbf{Z}^H\} = 1 - \frac{1}{2} \lVert \hat{\mathbf{D}}_q - \hat{\mathbf{Z}}^H \rVert^2_2
    \label{eq:svdphat_UqvZv}
\end{equation}
and thus maximizing (\ref{eq:svdphat_qmax2}) now corresponds to minimizing $\lVert \hat{\mathbf{D}}_q - \hat{\mathbf{Z}}^H \rVert^2_2$.
This minimization can be done by computing $\lVert \hat{\mathbf{D}}_q - \hat{\mathbf{Z}}^H \rVert^2_2$ for all values of $q$ and finding $q$ that leads to the minimum value, but this brings us back to the linear complexity order $\mathcal{O}(Q)$ as in (\ref{eq:svdphat_qmax2}).
Fortunately, the new formulation based on sum of squares becomes a nearest neighbor search problem, which can be solved efficiently using a k-d tree \cite{bentley1975multidimensional}.

Algorithm \ref{alg:svdphat_offline} summarizes the offline configuration performed prior to processing and the online computations.
The real-time performances are independent of the computationally expensive SVD and tree construction since these are done offline.
During online processing, computing the vector $\mathbf{Z}$ involves a complexity order of $\mathcal{O}(KPN)$ and the k-d tree search exhibits on average a complexity $\mathcal{O}(\log Q)$ \cite{bentley1975multidimensional}.
\begin{algorithm}[!ht]
    \vspace{4pt}
    \textbf{Offline:}
    \begin{algorithmic}[1]
        \STATE Generate $\mathbf{W}$ from (\ref{eq:srpphat_tdoa}), (\ref{eq:srpphat_W}), and (\ref{eq:svdphat_Wm}).
        \STATE Perform SVD and obtain $\mathbf{U}$, $\mathbf{S}$ and $\mathbf{V}$, with $K$ chosen according to condition in (\ref{eq:svdphat_trace}).
        \STATE Generate the normalized vectors $\hat{\mathbf{D}}_q$ from $\mathbf{D}_q$ in (\ref{eq:svdphat_Dv}).
        \STATE Build a k-d tree for all $\hat{\mathbf{D}}_q$.
    \end{algorithmic}
    \textbf{Online:}
    \begin{algorithmic}[1]
        \STATE Generate $\mathbf{X}$ from the STFT coefficients as in (\ref{eq:svdphat_Xv}).
        \STATE Compute $\mathbf{Z}$ with (\ref{eq:svdphat_Zv}) and generate $\hat{\mathbf{Z}}^H$.
        \STATE Find $\bar{q}$ using the k-d tree search.
        \STATE Find $Y_{\bar{q}}$ with the corresponding row of $\mathbf{W}$ in (\ref{eq:svdphat_Yv}).
    \end{algorithmic}    
    \caption{SVD-PHAT}
    \label{alg:svdphat_offline}
\end{algorithm}

\section{RESULTS}
\label{sec:results}

The parameters for the experiments are summarized in Table \ref{tab:results_parameters}.
The sample rate $f_S$ captures all the frequency content of speech, and the speed of sound $c$ corresponds to typical indoor conditions.
The frame size $N$ analyzes segments of $16$ msecs, and the hop size $\Delta N$ provides a $50\%$ overlap.
The potential DOA are represented by equidistant points on a unit sphere generated recursively from a tetrahedron, for a total of 2562 points, as in \cite{valin2007robust}.
\begin{table}[!ht]
    \centering
    \caption{SVD-PHAT Parameters}
    \vspace{5pt}
    \renewcommand{\arraystretch}{1.1}    
    \begin{tabular}{|ccccc|}
        \hline
        $f_S$ & $c$ & $N$ & $\Delta N$ & $Q$ \\
        \hline
        $16000$ & $343.0$ & $256$ & $128$ & $2562$\\
        \hline
    \end{tabular}
    \label{tab:results_parameters}
\end{table}

\begin{figure*}[!ht]
    \centering
    \subfloat[1-D linear array]{%
        \begin{tikzpicture}
        \begin{axis}[width=0.25\textwidth,xlabel=$x$,ylabel=$y$,zlabel=$z$,xtick={-1,0,1},ytick={-1,0,1},ztick={-5,0,5}]
            \addplot3[color=red,mark=*,only marks] 
            coordinates { (\aiMiCx,\aiMiCy,\aiMiCz)
                          (\aiMiiCx,\aiMiiCy,\aiMiiCz)
                          (\aiMiiiCx,\aiMiiiCy,\aiMiiiCz)
                          (\aiMivCx,\aiMivCy,\aiMivCz)
                          (\aiMvCx,\aiMvCy,\aiMvCz)
                          (\aiMviCx,\aiMviCy,\aiMviCz)
                          (\aiMviiCx,\aiMviiCy,\aiMviiCz) };
            \addplot3[color=red,mark=none]
            coordinates { (\aiMiCx,\aiMiCy,\aiMiCz) (\aiMviiCx,\aiMviiCy,\aiMviiCz) };
        \end{axis}
        \end{tikzpicture}
    }
    \hspace{10pt}
    \subfloat[2-D planar array]{%
        \begin{tikzpicture}
        \begin{axis}[width=0.25\textwidth,xlabel=$x$,ylabel=$y$,zlabel=$z$,xtick={-5,0,5},ytick={-4,0,4},ztick={-1,0,1}]
            \addplot3[color=blue,mark=*,only marks] 
            coordinates { (\aiiMiCx,\aiiMiCy,\aiiMiCz)
                          (\aiiMiiCx,\aiiMiiCy,\aiiMiiCz)
                          (\aiiMiiiCx,\aiiMiiiCy,\aiiMiiiCz)
                          (\aiiMivCx,\aiiMivCy,\aiiMivCz)
                          (\aiiMvCx,\aiiMvCy,\aiiMvCz)
                          (\aiiMviCx,\aiiMviCy,\aiiMviCz)
                          (\aiiMviiCx,\aiiMviiCy,\aiiMviiCz) };
            \addplot3[color=blue,mark=none]
            coordinates { (\aiiMiiCx,\aiiMiiCy,\aiiMiiCz)
                          (\aiiMiiiCx,\aiiMiiiCy,\aiiMiiiCz)
                          (\aiiMivCx,\aiiMivCy,\aiiMivCz)
                          (\aiiMvCx,\aiiMvCy,\aiiMvCz)
                          (\aiiMviCx,\aiiMviCy,\aiiMviCz)
                          (\aiiMviiCx,\aiiMviiCy,\aiiMviiCz)
                          (\aiiMiiCx,\aiiMiiCy,\aiiMiiCz) };
            \addplot3[color=blue,mark=none]
            coordinates { (\aiiMiiCx,\aiiMiiCy,\aiiMiiCz)
                          (\aiiMvCx,\aiiMvCy,\aiiMvCz) };                          
            \addplot3[color=blue,mark=none]
            coordinates { (\aiiMiiiCx,\aiiMiiiCy,\aiiMiiiCz)
                          (\aiiMviCx,\aiiMviCy,\aiiMviCz) };                          
            \addplot3[color=blue,mark=none]
            coordinates { (\aiiMivCx,\aiiMivCy,\aiiMivCz)
                          (\aiiMviiCx,\aiiMviiCy,\aiiMviiCz) };                          
        \end{axis}
        \end{tikzpicture}
    }
    \hspace{10pt}
    \subfloat[3-D array]{%
        \begin{tikzpicture}
        \begin{axis}[width=0.25\textwidth,xlabel=$x$,ylabel=$y$,zlabel=$z$]
            \addplot3[color=green,mark=*,only marks] 
            coordinates { (\aiiiMiCx,\aiiiMiCy,\aiiiMiCz)
                          (\aiiiMiiCx,\aiiiMiiCy,\aiiiMiiCz)
                          (\aiiiMiiiCx,\aiiiMiiiCy,\aiiiMiiiCz)
                          (\aiiiMivCx,\aiiiMivCy,\aiiiMivCz)
                          (\aiiiMvCx,\aiiiMvCy,\aiiiMvCz)
                          (\aiiiMviCx,\aiiiMviCy,\aiiiMviCz)
                          (\aiiiMviiCx,\aiiiMviiCy,\aiiiMviiCz) };
            \addplot3[color=green,mark=none] 
            coordinates { (\aiiiMiiCx,\aiiiMiiCy,\aiiiMiiCz)
                          (\aiiiMiiiCx,\aiiiMiiiCy,\aiiiMiiiCz) };                          
            \addplot3[color=green,mark=none] 
            coordinates { (\aiiiMivCx,\aiiiMivCy,\aiiiMivCz)
                          (\aiiiMvCx,\aiiiMvCy,\aiiiMvCz) };                          
            \addplot3[color=green,mark=none] 
            coordinates { (\aiiiMviCx,\aiiiMviCy,\aiiiMviCz)
                          (\aiiiMviiCx,\aiiiMviiCy,\aiiiMviiCz) };                          
            \addplot3[color=green,mark=none] 
            coordinates { (\aiiiMiiCx,\aiiiMiiCy,\aiiiMiiCz)
                          (\aiiiMivCx,\aiiiMivCy,\aiiiMivCz)
                          (\aiiiMiiiCx,\aiiiMiiiCy,\aiiiMiiiCz)
                          (\aiiiMvCx,\aiiiMvCy,\aiiiMvCz)
                          (\aiiiMiiCx,\aiiiMiiCy,\aiiiMiiCz) };
            \addplot3[color=green,mark=none] 
            coordinates { (\aiiiMiiCx,\aiiiMiiCy,\aiiiMiiCz)
                          (\aiiiMviCx,\aiiiMviCy,\aiiiMviCz)
                          (\aiiiMiiiCx,\aiiiMiiiCy,\aiiiMiiiCz)
                          (\aiiiMviiCx,\aiiiMviiCy,\aiiiMviiCz)
                          (\aiiiMiiCx,\aiiiMiiCy,\aiiiMiiCz) };                          
            \addplot3[color=green,mark=none] 
            coordinates { (\aiiiMivCx,\aiiiMivCy,\aiiiMivCz)
                          (\aiiiMviCx,\aiiiMviCy,\aiiiMviCz)
                          (\aiiiMvCx,\aiiiMvCy,\aiiiMvCz)
                          (\aiiiMviiCx,\aiiiMviiCy,\aiiiMviiCz)
                          (\aiiiMivCx,\aiiiMivCy,\aiiiMivCz) };           
        \end{axis}
        \end{tikzpicture}
    }    
    \caption{Geometries of the microphone arrays in xyz-coordinates (dimensions are given in cm)}
    \label{fig:mics}
\end{figure*}
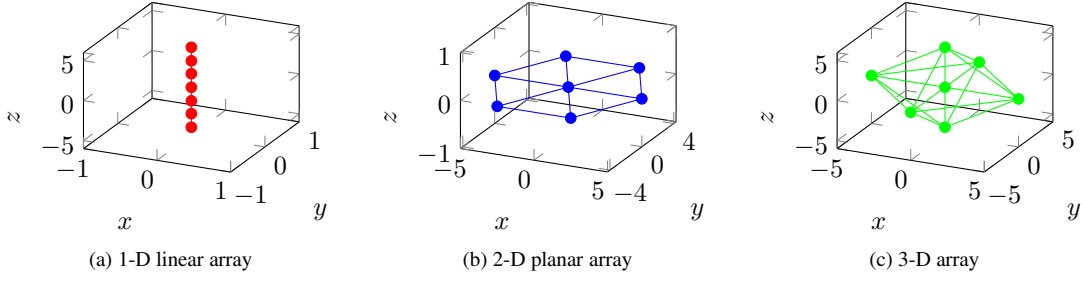

We investigate three different microphone array geometries: a 1-D linear array, a 2-D planar array and a 3-D array.
The microphones exact xyz-positions are given in cm in Table \ref{tab:results_positions} and the geometries are shown in Fig. \ref{fig:mics}.
\begin{table}[!ht]
    \centering
    \caption{Positions (x,y,z) of the microphones in cm}
    \vspace{5pt}
    \renewcommand{\arraystretch}{1.1}
    \begin{tabular}{|c|c|c|c|}
    \hline
    Mic & 1-D & 2-D & 3-D \\
    \hline
    $1$ & $(\aiMiCx,\aiMiCy,\aiMiCz)$ 
        & $(\aiiMiCx,\aiiMiCy,\aiiMiCz)$ 
        & $(\aiiiMiCx,\aiiiMiCy,\aiiiMiCz)$ \\
    $2$ & $(\aiMiiCx,\aiMiiCy,\aiMiiCz)$ 
        & $(\aiiMiiCx,\aiiMiiCy,\aiiMiiCz)$ 
        & $(\aiiiMiiCx,\aiiiMiiCy,\aiiiMiiCz)$ \\
    $3$ & $(\aiMiiiCx,\aiMiiiCy,\aiMiiiCz)$ 
        & $(\aiiMiiiCx,\aiiMiiiCy,\aiiMiiiCz)$ 
        & $(\aiiiMiiiCx,\aiiiMiiiCy,\aiiiMiiiCz)$ \\
    $4$ & $(\aiMivCx,\aiMivCy,\aiMivCz)$ 
        & $(\aiiMivCx,\aiiMivCy,\aiiMivCz)$ 
        & $(\aiiiMivCx,\aiiiMivCy,\aiiiMivCz)$ \\
    $5$ & $(\aiMvCx,\aiMvCy,\aiMvCz)$ 
        & $(\aiiMvCx,\aiiMvCy,\aiiMvCz)$ 
        & $(\aiiiMvCx,\aiiiMvCy,\aiiiMvCz)$ \\
    $6$ & $(\aiMviCx,\aiMviCy,\aiMviCz)$ 
        & $(\aiiMviCx,\aiiMviCy,\aiiMviCz)$ 
        & $(\aiiiMviCx,\aiiiMviCy,\aiiiMviCz)$ \\
    $7$ & $(\aiMviiCx,\aiMviiCy,\aiMviiCz)$
        & $(\aiiMviiCx,\aiiMviiCy,\aiiMviiCz)$ 
        & $(\aiiiMviiCx,\aiiiMviiCy,\aiiiMviiCz)$ \\
    \hline
    \end{tabular}
    \label{tab:results_positions}
\end{table}

Simulations are conducted to measure the accuracy of the proposed method.
The microphone array and the target source are positionned randomly in a 10m x 10m x 3m rectangular room.
For each configuration, the room reverberation is modeled with Room Impulse Responses (RIRs) generated with the image method \cite{allen1979image}, where the reflection coefficients are sampled randomly in the uniform interval between 0.2 and 0.5.
Sound segments from the TIMIT dataset are then convolved with the generated RIRs.
Diffuse white noise is added on each channel, for a signal-to-noise ratio (SNR) that varies randomly between 0dB and 30dB.
A total of 1000 different configurations are generated for each microphone array.

We vary $\delta$ to analyze its impact on the accuracy of localization, measured as Root Mean Square Error (RMSE).
For the 1-D linear array, localization can only provide a position on $180^{\circ}$ arc.
The 3-D position from the 2562 points unit sphere is therefore mapped to an arc:
\begin{equation}
    f_1(\mathbf{s}) = [\cos(g(\mathbf{s})), 0, \sin(g(\mathbf{s}))]
\end{equation}
where
\begin{equation}
    g(\mathbf{s}) = \atantwo\left\{\mathbf{s}|_z,\sqrt{(\mathbf{s}|_x)^2+(\mathbf{s}|_y)^2}\right\}
\end{equation}

For the 2-D planar array, localization returns a position on a half-sphere, and thus every point is mapped to the positive hemisphere as follows:
\begin{equation}
    f_2(\mathbf{s}) = [\mathbf{s}|_x, \mathbf{s}|_y, |\mathbf{s}|_z|]
\end{equation}

Finally, for the 3-D array, the localization result can span the full 3-D space, such that the mapping function corresponds to identity:
\begin{equation}
    f_3(\mathbf{s}) = \mathbf{s}
\end{equation}

The RMSE between the estimated DOA ($\mathbf{s}_{\bar{q}}$) for all frames for a given room configuration and speech signal is summed and weighted with the energy $Y_q$, and then compared with the theoretical DOA defined by $\mathbf{s}_0$.
The mathematical expression corresponds to (\ref{eq:results_rmse}), where $\alpha = \{1,2,3\}$ for 1-D, 2-D and 3-D arrays, respectively.
\begin{equation}
    \textrm{RMSE}_{\alpha} = \left\lVert\frac{\sum f_{\alpha}(\mathbf{s}_{\bar{q}})Y_{\bar{q}}}{\sum Y_{\bar{q}}} - f_{\alpha}(\mathbf{s}_0)\right\rVert_2
    \label{eq:results_rmse}
\end{equation}

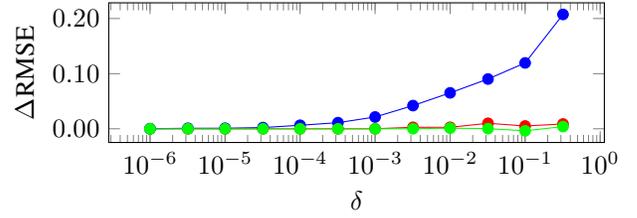
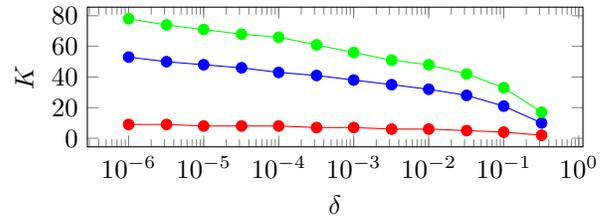
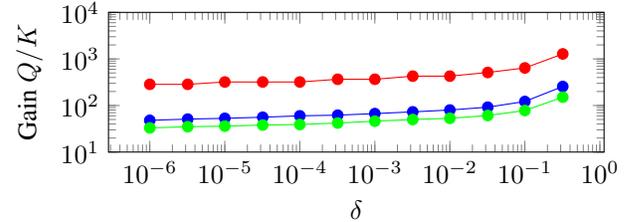
\begin{figure}[!ht]
    \centering
    \subfloat[Difference between the Root Mean Square Error of SVD-PHAT and the exact SRP-PHAT -- smaller is better.]{%
        \begin{tikzpicture}
        \begin{axis}[xmode=log,
                     xlabel=$\delta$,
                     ylabel=$\Delta$RMSE,
                     y tick label style={/pgf/number format/.cd,fixed,fixed zerofill,precision=2,/tikz/.cd},
                     height=0.4\columnwidth,
                     width=0.95\columnwidth]
        \addplot[red,mark=*] table [x=delta, y=rmse, col sep=comma] {data/rmse_delta_1d.csv};
        \addplot[blue,mark=*] table [x=delta, y=rmse, col sep=comma] {data/rmse_delta_2d.csv};        
        \addplot[green,mark=*] table [x=delta, y=rmse, col sep=comma] {data/rmse_delta_3d.csv};
        \end{axis}
        \end{tikzpicture}
        \label{fig:results_rmse}
    }\\
    \vspace{10pt}
    \hspace{5pt}
    \subfloat[Value of the variable $K$ (the rank of the decomposition) for the proposed SVD-PHAT method -- smaller is better.]{%
        \begin{tikzpicture}
        \begin{axis}[xmode=log,xlabel=$\delta$,ylabel=$K$,height=0.4\columnwidth,width=0.95\columnwidth]
        \addplot[red,mark=*] table [x=delta, y=K, col sep=comma] {data/rmse_delta_1d.csv};
        \addplot[blue,mark=*] table [x=delta, y=K, col sep=comma] {data/rmse_delta_2d.csv};        
        \addplot[green,mark=*] table [x=delta, y=K, col sep=comma] {data/rmse_delta_3d.csv};
        \end{axis}
        \end{tikzpicture}  
        \label{fig:results_K}
    }\\
    \subfloat[Performance gain of SVD-PHAT when compared to exact SRP-PHAT method -- greater is better.]{%
        \begin{tikzpicture}
        \begin{axis}[xmode=log,ymode=log,xlabel=$\delta$,ylabel=Gain $Q/K$,ymin=10,ymax=10000,height=0.4\columnwidth,width=0.95\columnwidth]
        \addplot[red,mark=*] table [x=delta, y=gain, col sep=comma] {data/rmse_delta_1d.csv};
        \addplot[blue,mark=*] table [x=delta, y=gain, col sep=comma] {data/rmse_delta_2d.csv};        
        \addplot[green,mark=*] table [x=delta, y=gain, col sep=comma] {data/rmse_delta_3d.csv};
        \end{axis}
        \end{tikzpicture}  
        \label{fig:results_gain}
    }
    \caption{Performance of the proposed SVD-PHAT method with respect to the exact SRP-PHAT method. Results are presented for the 1-D linear array (red), the 2-D planar array (blue) and the 3-D array (green).}
    \label{fig:results}
\end{figure}

Figure \ref{fig:results_rmse} shows the difference between the RMSE from SVD-PHAT and SRP-PHAT (denoted as $\Delta$RMSE), with respect to the $\delta$ parameter.
As expected, when $\delta$ increases, the reconstruction error gets significant and this reduces the accuracy of localization for SVD-PHAT.
It is interesting to note that the 2-D planar array shows the largest increase in RMSE.
Figure \ref{fig:results_K} shows the value of the $K$ parameter as a function of $\delta$.
Note how the $K$ value is smaller when the array spans only one or two dimensions, as expected since the transfer function between DOAs are more correlated.
The gain in performance is mostly due to the reduction from a matrix multiplication with $Q$ rows in (\ref{eq:svdphat_Yv}) to a matrix multiplication with $K$ rows in (\ref{eq:svdphat_Zv}).
Figure \ref{fig:results_gain} therefore shows the gain $Q/K$ as a function of $\delta$.

It is reasonable to define $\delta = 10^{-5}$ as the RMSE between SRP-PHAT and SVD-PHAT is almost identical.
With this configuration, the gain $Q/K$ reaches $320$, $53$ and $36$ for 1-D, 2-D and 3-D arrays, which is considerable, and demonstrates the superiority of SVD-PHAT over SRP-PHAT in terms of computational requirements, while preserving the same accuracy.

\section{CONCLUSION}
\label{sec:conclusion}

This paper introduces a new localization method named SVD-PHAT.
This technique can perform with the same accuracy as SRP-PHAT, while reducing significantly the amount of computations.

In future work, we will investigate multiple source localization with SVD-PHAT.
It would also be interesting to introduce binary time-frequency mask, which could reduce even more the amount of computations.
The method could also be extended to deal with speed of sound mismatch, the near-field effect and microphone position uncertainty.

\vfill\pagebreak

\bibliographystyle{IEEEbib}
\bibliography{refs}

\end{document}